\documentclass[eqsecnum,aps,preprint]{revtex4}
\usepackage{epsfig,graphicx}
\newcommand{\T}[1]{\text{T} \left[ #1 \right]}
\newcommand{\St}[1]{| #1 \rangle}
\newcommand{\sT}[1]{\langle #1 |}
\newcommand{\Gr}{\St{\Omega}}
\newcommand{\gR}{\sT{\Omega}}
\newcommand{\GGVren}{\langle \hspace{-0pt} \frac{\alpha_s}{\pi} G^2 \hspace{-0pt} \rangle}
\newcommand{\GGMren}{\langle \hspace{-0pt} \frac{\alpha_s}{\pi} \hspace{-2pt} \left( \frac{(vG)^2}{v^2} - \frac{G^2}{4} \right) \hspace{-0pt} \rangle}
\DeclareSymbolFont{rsfs}{U}{rsfs}{m}{n}
\DeclareSymbolFontAlphabet{\mathrsfs}{rsfs}
\usepackage{amsmath}
\usepackage{amssymb}
\usepackage{nicefrac}

\begin{document}

\title{QCD sum rules for D and B mesons in nuclear matter}

\author{{\sc T.\ Hilger, R.\ Thomas, B.\ K\"ampfer}}

\affiliation{
Forschungszentrum Dresden-Rossendorf, PF 510119, D-01314 Dresden, Germany\\
TU Dresden, Institut f\"ur Theoretische Physik, 01062 Dresden, Germany}

\begin{abstract}
QCD sum rules for $D$ and $B$ mesons embedded in cold nuclear matter are evaluated. 
We quantify the mass splitting of $D - \overline D$ and $B - \overline B$ mesons as a function of the nuclear matter density; extrapolated to saturation density it is in the order of $60$ and $130$ MeV, respectively, driven essentially by the condensates $\langle q^\dagger q \rangle$, $\langle q^\dagger g \sigma \mathrsfs{G} q \rangle$ and $\langle \overline{q} q \rangle$. 
The genuine chiral condensate $\langle \overline{q} q \rangle$, amplified by heavy-quark masses, enters the Borel transformed sum rules for the mass splitting beyond linear density dependence.
Including strange quark condensates reveals a numerically smaller and opposite effect for the $D_s - \overline D_s$ mass splitting.
\end{abstract}

\maketitle

\section{Introduction}
\label{sct:introduction}

QCD sum rules offer a link from hadronic properties, encoded in spectral functions, 
to QCD related quantities, like condensates, in the non-perturbative domain. 
A particularly valuable aspect of QCD sum rules is, therefore, the possibility 
to predict in-medium modifications of hadrons, supposed the density and 
temperature dependence of the relevant condensates is known. 
Taking the attitude that this is the case, one arrives at testable predictions 
for changes of hadronic properties in an ambient strongly interacting medium. 
There is a vast amount of literature on the in-medium changes of light vector mesons, 
cf.\ \cite{Hatsuda_Lee,Leupold_Mosel,Ruppert,Mosel,Mosel2,Erice,our_PRL,Rapp_Wambach,Lutz} and further references therein.
Vector mesons are of interest as their spectral functions determine, e.g., 
the dilepton emissivity of hot and compressed nuclear matter. 
Via the direct decays $V \to l^+ l^-$, where $V$ stands for a vector meson 
and $l^+ l^-$ for a dilepton, the spectral distribution of $V$ can be probed experimentally. 
Accordingly, heavy-ion experiments are often accompanied by special devices for measurements 
of $l^+ l^- = e^ + e^-$ or $\mu^+ \mu^-$. 
Addressed questions concern in particular signals for chiral restoration 
\cite{Rapp_Wambach}. Clearly, besides the QCD sum rules, 
also purely hadronic models have been employed to understand the behavior of vector mesons 
in nuclear matter, cf. \cite{Mosel,Rapp_Wambach,Lutz} for examples.

Such hadronic models are also used in the strangeness sector \cite{Weise_Kpm}. Here, the distinct behavior of kaons and anti-kaons attracted much attention, cf.\ \cite{Scheinast} for experimental aspects. The upcoming accelerator complex FAIR at GSI/Darmstadt offers the opportunity to extend the experimental studies into the charm sector. The CBM collaboration \cite{CBM} intends to study the near-threshold production of $D$ and $J/\psi$ mesons in heavy-ion-collisions, while the PANDA collaboration \cite{PANDA} will focus on charm spectroscopy, as well as on charmed mesons produced by anti-proton annihilation in nuclei.
In the CBM experiments, charm degrees of freedom will serve as probes of nuclear matter at the maximum compression achievable in the laboratory, at moderate temperatures. Despite of this interest in $D$ mesons and their behavior in nuclear matter, the literature on in-medium $D$ mesons is fairly scarce. While there is a variety of calculations within a hadronic basis, e.g.\ \cite{Tolos,Angel, Lutz_charm}, or within the quark-meson coupling model, e.g.\ \cite{Tsushima}, the use of QCD sum rules is fairly seldom \cite{Hayashigaki,Morath,D_Zschocke}. In contrast, the treatment of vacuum $D$ (and $D_s$) ground states is performed in a concise manner \cite{Hayashigaki_Terasaki,Aliev,Bengt_Student}.

The aim of the present paper is the re-evaluation of the QCD sum rules for $D$ and $\overline D$ mesons in cold nuclear matter and an extension to $B$ and $\overline B$ mesons as well. Even for the operator product expansion (OPE) up to mass dimension 5, there are conflicting results in the literature concerning the open charm sector \cite{Hayashigaki,Morath,D_Zschocke,Aliev,Neubert,Narison1,Narison2}.
While in \cite{Hayashigaki} only the even part of the in-medium OPE up to mass dimension 4 has been used, we present here the even as well as the odd in-medium OPE up to mass dimension 5. Moreover, a term $\propto \langle\overline q g \sigma \mathrsfs{G} q \rangle$, i.e.\ the lowest-order quark-gluon condensate, can be found in the literature with various factors and signs already for the vacuum. As the subtle $D - \overline D$ mass splitting is of paramount experimental interest, a safe basis is mandatory.

Our article is organized as follows. Section \ref{sct:qcd_sum_rules} contains the QCD sum rules formalism for $D$ and $\overline D$ mesons. The spectral functions are discussed in section \ref{sct:parameterizing_the_spectral_function}. The numerical evaluation for $D, \overline D$ and $B, \overline B$ mesons is executed in sections \ref{sct:evaluation_for_d_dbar_mesons} and \ref{sct:evaluation_for_b_bbar_mesons}. The discussion and summary can be found in section \ref{sct:summary}.

%================================================
\section{QCD sum rules}
\label{sct:qcd_sum_rules}

The basic quantity to be evaluated is the two-point function
\begin{equation}
\Pi(q) = i \int d^4x \, e^{iqx} \gR \T{\text{j}(x) \text{j}^\dagger(0)} \Gr
\label{correlator}
\end{equation}
as the Fourier transform of the expectation value of the time-ordered product of the currents $\text{j}(x)$ and $\text{j}^\dagger(0)$; the state $\vert \Omega \rangle$ has properties $\mathrsfs{H} \Gr = E_\Omega \Gr , \: \langle \Omega | \Omega \rangle = 1 , \: a \Gr \neq 0$. $\mathrsfs{H}$ is the full Hamiltonian of the theory, $a$ an arbitrary annihilation operator, and the field operators are taken in the Heisenberg picture. Splitting up $\Pi (q_0,\vec{q}\,)$ into an even ($e$) and an odd ($o$) part according to $\Pi(q_0,\vec{q}\,) = \Pi^e(q^2_0,\vec{q}\,) + q_0 \Pi^o(q^2_0,\vec{q}\,) $ with
\begin{subequations}
\begin{flalign}
    \Pi^e(q_0,\vec{q}\,) &= \frac{1}{2} \left( \Pi(q_0,\vec{q}\,) + \Pi(-q_0,\vec{q}\,)\right)
        = \Pi^e(-q_0,\vec{q}\,) \: ,
    \label{eq:def_even}
    \\
    \Pi^o(q_0,\vec{q}\,) &= \frac{1}{2q_0} \left( \Pi(q_0,\vec{q}\,) - \Pi(-q_0,\vec{q}\,)\right)
        = \Pi^o(-q_0,\vec{q}\,) \: ,
    \label{eq:def_odd}
\end{flalign}
\end{subequations}
one arrives at the $N$-fold subtracted dispersion relations in the complex $q_0$ plane
\begin{subequations} \label{eq:disp_rel}
\begin{equation}
\begin{split}
    & \Pi^e(q_0,\vec{q}\,) - \frac{1}{2} \sum_{n=0}^{N-1} \frac{\Pi^{(n)}(0,\vec{q}\,)}{n!} (q_0)^n \left(1+(-1)^n\right)
    \\& \quad
    = \frac{1}{2\pi} \int_{-\infty}^{+\infty} ds \, \Delta \Pi(s,\vec{q}\,) \frac{q_0^N}{s^{N-1}}
        \frac{\left(1+(-1)^N\right) + \frac{q_0}{s}\left(1-(-1)^N\right)}{s^2-q_0^2},
\label{eq:even_disp_rel}
\end{split}
\end{equation}
\begin{equation}
\begin{split}
    & \Pi^o(q_0,\vec{q}\,)
    - \frac{1}{2} \sum_{n=0}^{N-1} \frac{\Pi^{(n)}(0,\vec{q}\,)}{n!} (q_0)^{n-1} \left(1-(-1)^n\right)
    \\& \quad
    = \frac{1}{2\pi} \int_{-\infty}^{+\infty} ds \, \Delta \Pi(s,\vec{q}\,) \frac{q_0^{N-1}}{s^{N-1}}
        \frac{\left(1-(-1)^N\right) + \frac{q_0}{s}\left(1+(-1)^N\right)}{s^2-q_0^2}
\label{eq:odd_disp_rel}
\end{split}
\end{equation}
\end{subequations}
with $\Delta \Pi(s) = {\rm Im} \Pi(s)$. From Eq.~\eqref{eq:disp_rel} we see that $\Pi^{e,o}$ depend on $q_0^2$. The Borel transformed sum rules are
\begin{subequations} \label{eq:b_disp}
\begin{equation}
\begin{split}
    & \mathrsfs{B} \left[ \Pi^e_{OPE}(\omega^2,\vec{q}\,) \right] \left( M^2 \right) 
    \\& \quad
    = \left[ \frac{1}{\pi} \int_{s_0^-}^{s_0^+} ds \,
        \Delta \Pi(s,\vec{q}\,)
        + \frac{1}{\pi} \left( \int_{-\infty}^{s_0^-} + \int^{+\infty}_{s_0^+} \right) ds \,
        \Delta \Pi(s,\vec{q}\,) \right]
        s e^{-s^2/M^2}
    \: ,
\label{eq:even_b_sr}
\end{split}
\end{equation}
\begin{equation}
\begin{split}
    & \mathrsfs{B} \left[ \Pi^o_{OPE}(\omega^2,\vec{q}\,) \right] \left( M^2 \right)
    \\& \quad
    = \left[ \frac{1}{\pi} \int_{s_0^-}^{s_0^+} ds \, \Delta \Pi(s,\vec{q}\,)
        + \frac{1}{\pi} \left( \int_{-\infty}^{s_0^-} + \int^{+\infty}_{s_0^+} \right) ds \,
        \Delta \Pi(s,\vec{q}\,) \right]
        e^{-s^2/M^2} \: ,
\label{eq:odd_b_sr}
\end{split}
\end{equation}
\end{subequations}
where the subscript OPE denotes the operator product expansion of $\gR \T{\text{j}(x) \text{j}^\dagger(y)} \Gr = \sum_{O} C_O(x-y) \gR O \Gr$ with QCD condensates $ \gR O \vert\Omega \rangle$ and Wilson coefficients $C_O$. We are interested in the low-lying strength encoded in  $\int_{s_0^-}^{s_0^+} ds \, \Delta \Pi s e^{-s^2/M^2}$ and $\int_{s_0^-}^{s_0^+} ds \, \Delta \Pi e^{-s^2/M^2}$, while the continuum parts $\left( \int_{-\infty}^{s_0^-} + \int^{+\infty}_{s_0^+} \right) ds \, \Delta \Pi s e^{-s^2/M^2}$ and $\left( \int_{-\infty}^{s_0^-} + \int^{+\infty}_{s_0^+} \right) ds \, \Delta \Pi e^{-s^2/M^2}$ will be merged into the perturbative OPE part $\Pi_{D^+}^{per}(s)$ (see below) according to the semi-local duality hypothesis; $s^\pm_0$ are the corresponding continuum thresholds; $M$ is the Borel mass.

Employing the current operator $\text{j}_{D^+} = i \overline{d} \gamma_5 c$ (and $\text{j}_{D^-} = \text{j}^\dagger_{D^+}(x) = i \overline{c} \gamma_5 d$ for the antiparticle), we obtain for the OPE side up to mass dimension 5, in the rest frame of nuclear matter $v = (1, \vec{0} \,)$ ($v$ stands for the medium four-velocity), in the limit $m_d \to 0$ and sufficiently large charm-quark pole mass $m_c$,
\begin{subequations} \label{eq:borel_sr_d}
\begin{align}
    & \mathrsfs{B} \left[ \Pi^e_{OPE}(\omega^2, \vec{q} = 0\,) \right] \left( M^2 \right)
    \nonumber
    \\& \quad =
        \frac{1}{\pi} \int_{m_c^2}^\infty ds e^{-s/M^2} \text{Im} \Pi_{D^+}^{per}(s, \vec{q} = 0\,)
        \nonumber
        \\& \quad
        + e^{-m_c^2/M^2} \left(
        -m_c \langle \overline{d}d \rangle
        + \frac{1}{2} \left( \frac{m_c^3}{2M^4} - \frac{m_c}{M^2} \right)
        \langle \overline{d}g\sigma\mathrsfs{G}d \rangle
        + \frac{1}{12} \GGVren \right.
        \nonumber
        \\& \quad
        + \left[ \left( \frac{7}{18} + \frac{1}{3} \ln \frac{\mu^2 m_c^2}{M^4}
         - \frac{2\gamma_E}{3} \right)
        \left( \frac{m_c^2}{M^2} - 1 \right) - \frac{2}{3} \frac{m_c^2}{M^2}
        \right] \GGMren
        \nonumber
        \\& \quad \left.
        + 2 \left( \frac{m_c^2}{M^2} - 1 \right) \langle d^\dagger iD_0 d \rangle
        + 4 \left( \frac{m_c^3}{2M^4} - \frac{m_c}{M^2} \right)
        \left[ \langle \overline{d} D_0^2 d \rangle
        - \frac{1}{8} \langle \overline{d} g \sigma \mathrsfs{G} d \rangle \right]
        \right) \: ,
\label{eq:borel_sr_d_even}
        \displaybreak[0]
    \\
    & \mathrsfs{B} \left[ \Pi^o_{OPE}(\omega^2, \vec{q} = 0\,) \right] \left( M^2 \right)
            \nonumber
        \\& \quad =
        e^{-m_c^2/M^2} \left(
        \langle d^\dagger d \rangle
        - 4 \left( \frac{m_c^2}{2M^4} - \frac{1}{M^2} \right) \langle d^\dagger D_0^2 d \rangle
        - \frac{1}{M^2} \langle d^\dagger g \sigma \mathrsfs{G} d \rangle
        \right)
        \: ,
\label{eq:borel_sr_d_odd}
\end{align}
\end{subequations}
where $\alpha_s = g^2 /4 \pi$. (Analog relations hold for $\text{j}_{D^0}(x) = i \overline{u} \gamma_5 c$ with $\text{j}_{\overline{D}^0} (x) = \text{j}^\dagger_{D^0}(x) = i \overline{c} \gamma_5 u$.) The calculational details are documented in \cite{Hilger}. While the perturbative spectral function $\text{Im} \Pi_{D^+}^{per}(s)$ (see \cite{Aliev,Narison1} for an explicit representation in terms of the pole mass) is known for a long time, discrepancies especially for Wilson coefficients of medium specific condensates exist. An important intermediate step is the careful consideration of the operator mixing, which occurs due to the introduction of non-normal ordered condensates and the corresponding cancellation of infrared divergent terms $\propto m_q^{-2}$ and $\log m_q$ ($m_q$ is the light-quark mass) at zero and non-zero densities \cite{Hilger}. This is not to be confused with the operator mixing within renormalization group methods.
In vacuum our expression differs from \cite{Hayashigaki} in the prefactor of $\langle (\alpha_s/\pi) G^2 \rangle$; \cite{Aliev} reports an opposite sign; \cite{Morath} finds the same result. For the medium case \cite{Morath} does not give explicit results, while terms $\propto \langle \overline{d} d \rangle$, $\propto \langle (\alpha_s/\pi) G^2 \rangle$, $\propto \langle (\alpha_s/\pi) ((vG)^2/v^2 - G^2/4)\rangle$ have different prefactors compared to \cite{Hayashigaki}. Higher order terms are partially considered in \cite{Morath} and are found to be numerically not important.

We stress the occurrence of the term $m_c \langle \overline{d}d \rangle$. In the pure light quark sector, say for vector mesons, it would read $m_d \langle \overline{d}d \rangle$, i.e., the small down-quark mass strongly suppresses the numerical impact of the chiral condensate $\langle \overline{d}d \rangle$. In fact, only within the doubtful factorization of four-quark condensates into the squared chiral condensate it would become important \cite{our_PRL}. Here, the large charm-quark mass acts as an amplifier of the genuine chiral condensate entering the QCD sum rules for the $D^+$ meson.

%================================================
\section{Parameterizing the spectral function}
\label{sct:parameterizing_the_spectral_function}

Especially in vacuum the spectral strength of the iso-scalar--vector excitation exhibits a well-defined sharp peak (the $\omega$ meson) and a well-separated flat continuum. Assuming the same features for the $\omega$ meson in a medium gives rise to the often exploited "pole + continuum" ansatz. One way to avoid partially such a strong assumption is to introduce certain moments of the spectral function, thus replacing the assumed pole mass by a centroid of the distribution \cite{our_PRL,Weise}.

For $D$ mesons the sum rule includes an integral which arises from the dispersion relation over positive and negative energies, see Eq.~\eqref{eq:b_disp}. Similar to baryons \cite{Furnstahl,our NPA}, one may try to suppress the antiparticle contribution corresponding here to $D^-$. This, however, is not completely possible \cite{Erice}. Nevertheless, one can identify with the ansatz $\Delta \Pi (s) = \pi F_+ \, \delta (s - m_+) - \pi F_- \, \delta(s + m_-)$, motivated by the Lehmann-representation of the correlation function, the meaning of the even and odd sum rules \eqref{eq:b_disp} with \eqref{eq:borel_sr_d}:
\begin{subequations} \label{eq:mom_sys}
\begin{align}
e \equiv
\int_{s_0^-}^{s_0^+} ds \, s \, \Delta \Pi {\rm e}^{-s^2 /M^2} &= m_+ F_+ {\rm e}^{-m_+^2 / M^2}
                        + m_- F_- {\rm e}^{-m_-^2 / M^2}, \label{mom1}\\
o \equiv
\int_{s_0^-}^{s_0^+} ds \, \Delta \Pi {\rm e}^{-s^2 /M^2} &= F_+ {\rm e}^{-m_+^2 / M^2} -
                        F_- {\rm e}^{-m_-^2 / M^2} \label{mom2}.                       
\end{align}
\end{subequations}
With the decomposition $m_\pm = m \pm \Delta m$ and $F_\pm = F \pm \Delta F$ the leading order terms of an expansion in $\Delta m$ for the first and second lines become $\propto F m {\rm e}^{-m^2 /M^2}$ and $\propto (\Delta F  - 2 \Delta m \, F \frac{m}{M^2}) {\rm e}^{-m^2 /M^2}$ meaning that (\ref{mom1}) is related to the average $D + \overline D$ properties, while (\ref{mom2}) refers to the $D - \overline D$ splitting. If one assumes for the moment being $m_\pm$ and $F_\pm$ to be independent of the Borel mass $M$, \eqref{eq:mom_sys} can be rewritten as
\begin{subequations} \label{eq:m_sys}
\begin{align}
    \Delta m &= \frac12 \frac{ o e^\prime - e o^\prime}{ e^2 + o o^\prime } \: ,
    \label{eq:delta_m}
    \\
    m &= \sqrt{ \Delta m^2 - \frac{ e e^\prime + \left( o^\prime \, \right)^2 }{ e^2 + o o^\prime } } \: ,
    \label{eq:m_cms}
\end{align}
\end{subequations}
where a prime denotes the derivative w.r.t.\ $1/M^2$. In order to gain further insight into the dependencies of $\Delta m$ and $m$ on the different OPE contributions, we expand \eqref{eq:m_sys} up to first order in the density $n$ employing $e(n) \approx e(0) + n \, d e/ dn |_{n=0}$ and $o(n) \approx n \, do/dn|_{n=0}$, since $o(0)$ must vanish to reproduce the vacuum sum rules where $\Delta m(n=0) = 0$ holds. We remark that these expansions are exact for a linear density dependence of the condensates and if $s_0^2 = ( (s_0^{+})^{2} + (s_0^{-})^{2} ) / 2$ as well as $\Delta s_0^2 = ( (s_0^{+})^{2} - (s_0^{-})^{2} ) / 2$ are density independent and the Borel mass $M$ is kept fixed. This implies $\Delta s_0^2 = 0$ for all densities, because otherwise $o(0) = 0$ cannot be fulfilled. For small densities we get accordingly
\begin{subequations} \label{eq:small_dens_appr1}
\begin{align}
    \Delta m(n) & \approx \frac12 \frac{ \left. \frac{ { d} o }{ { dn} } \right|_{0}
        e^\prime(0) - e(0) \bigl. \frac{ { d} o^\prime }{ { dn} } \bigr|_{0}}{ e(0)^2 }\, n
    \: ,
    \\
    m(n) & \approx \sqrt{ - \frac{e^\prime(0)}{e(0)} }
        + \frac12 \sqrt{ - \frac{e(0)}{e^\prime(0)} }
        \frac{ \left. \frac{ { d} e }{ { dn} } \right|_{0} e^\prime(0)
        - e(0) \bigl. \frac{ { d} e^\prime }{ { dn} } \bigr|_{0}}{ e(0)^2 } \,n \: ,
\end{align}
\end{subequations}
which can be written as
\begin{subequations} \label{eq:small_dens_appr}
\begin{align}
    \Delta m(n) & \approx - \frac12
    \frac{ \left. \frac{ { d} o }{ { dn} } \right|_{0} m^2(0) + \bigl. \frac{ { d} o^\prime }{ { dn} } \bigr|_{0}}
        {e(0)} n
    \: ,
    \label{eq:small_dens_appr_1}
    \\
    m(n) & \approx m(0)
        - \frac{1}{2 m(0)}
        \frac{ \left. \frac{ { d} e }{ { dn} } \right|_{0} m^2(0)
        + \bigl. \frac{ { d} e^\prime }{ { dn} } \bigr|_{0} }{ e(0) } n \: .
        \label{eq:small_dens_appr_2}
\end{align}
\end{subequations}

Eq.~\eqref{eq:m_sys} and the approximations in \eqref{eq:small_dens_appr} offer a transparent interpretation. In vacuum ($n=0$), there is no mass splitting, of course; the mass parameter $m(0)$ is determined by the even part of the OPE. In first order of $n$, the mass splitting $\Delta m$ depends on both the even and odd parts of the OPE, whereas only the even part of the OPE determines the mass parameter $m$, having the meaning of the centroid of the doublet $D^+$, $D^-$.
If one is only interested in the mass shift of the iso-doublet as a whole, for small densities it is sufficient to consider the even OPE part alone, as was done in \cite{Hayashigaki}. However, for the mass splitting the odd part of the OPE is of paramount importance. In particular, it is the density dependence of the odd part of the OPE alone which drives the mass splitting in first order of $n$. Interestingly, the density dependent part of the chiral condensate, which belongs to the even part of the OPE, enters the mass splitting in order $n^2$. The chiral condensate comes about in the combination $m_c \langle \overline d d \rangle$. The large charm mass amplifies the numerical impact, as stressed above.

Furthermore, up to order $n$, only $s_0^2(n=0)$ and $\left. \frac{d \Delta s_0^2}{ dn } \right|_0$ enter $\Delta m$ (i.e.\ neither $\left. \frac{d s_0^2}{ dn } \right|_0$ nor $\left. \frac{d M}{ dn } \right|_0$), whereas $s_0^2(n=0)$, $\left. \frac{d s_0^2}{ dn } \right|_0$ and $\left. \frac{d M}{ dn } \right|_0$ enter $m$ (not $\left. \frac{d \Delta s_0^2}{ dn } \right|_0$) as can be seen from the derivatives needed to calculate $m$ and $\Delta m$ from \eqref{eq:small_dens_appr1}:
\begin{subequations} \label{eq:ntlo_terms}
\begin{align}
    \left. \frac{ { d} o }{ { dn} } \right|_{0} =&
        \left( \frac{ e^{-s_0^2/M^2}}{\pi s_0} \text{Im} \Pi_{per}(s_0^2) \frac{d \Delta s_0^2}{dn} \right)_{n=0}
        + \: \text{non-perturbative terms} \: ,
    \\ \nonumber
    \left. \frac{ { d} e }{ { dn} } \right|_{0} =&
        \left( \frac{ e^{-s_0^2/M^2}}{\pi} \text{Im} \Pi_{per}(s_0^2) \frac{d s_0^2}{dn}
        - \frac1\pi \int_{m_c^2}^{s_0^2} ds \text{Im} \Pi_{per}(s) s e^{-s/M^2} \frac{d M^{-2}}{dn} \right)_{n=0}
        \\ &
        + \: \text{non-perturbative terms} \: .
\end{align}
\end{subequations}

While \eqref{eq:small_dens_appr} suggests that one can independently adjust $m(0)$ to the respective vacuum value, Eq.~\eqref{eq:ntlo_terms} evidences that further vacuum parameters (such as $M$, $\left. \frac{d M}{ dn } \right|_0$, $s_0^2$, $\Delta s_0^2$, $\left. \frac{d s_0^2}{ dn } \right|_0$ and $\left. \frac{d \Delta s_0^2}{ dn } \right|_0$) enter the density dependence and have to be chosen consistently to the vacuum mass. That means, one has to evaluate the complete sum rule, including consistently the vacuum limit.

We remark that \eqref{eq:m_sys} or \eqref{eq:small_dens_appr} are a consequence of using a pole-ansatz for the first excitation. The OPE and the special form of the continuum contribution to the spectral integral are encoded in $e$ and $o$. Likewise, the arguments following \eqref{eq:m_sys} merely use $o(0) = 0$. The last point must always be fulfilled in any sum rule and/or dispersion relation, because at zero density, the current-current correlation function \eqref{correlator} only depends on $q^2$ and, hence, the odd part \eqref{eq:def_odd} vanishes. This can also be confirmed directly from \eqref{mom2}, where $s_0^+ = s_0^-$, due to particle anti-particle symmetry, and $\Delta \Pi (s) = \Delta \Pi(s^2)$, meaning that the spectral density in vacuum merely depends on the squared energy, on account for $o(0) = 0$.

To arrive at a more general result, one may seek for a relation of $m_\pm$ to certain normalized moments of $\Pi(s)$ 
(or ratios thereof)
independent of a special ansatz, as can be done
in the case of vector mesons \cite{our_PRL,Weise}. In this spirit one would be
tempted to define 
$\int_{0}^{s_0^+} ds \, s \Delta \Pi {\rm e}^{-s^2 /M^2} \to m_+ F_+ {\rm e}^{-m_+^2 / M^2}$
and $\int_{0}^{s_0^+} ds \, \Delta \Pi {\rm e}^{-s^2 /M^2} \to F_+ {\rm e}^{-m_+^2 / M^2}$
and analogously for $m_-$ and $F_-$.
However, such a separation of positive and negative frequency parts leads
to multiple but different expressions for $m_\pm$ which can be fulfilled consistently
only for special cases of $\Pi(s)$, as for the above pole ansatz. (This can 
be seen by combining these relations with derivatives according to $M^{-2}$.)
Therefore, one is left with either the somewhat vague statement that
\eqref{eq:mom_sys} refers to $D + \overline D$ and $D - \overline D$ properties
or one has to employ another explicit ansatz for the function $\Pi (s)$.

Alternatively, one can define moments which correspond to the integrals 
in \eqref{eq:mom_sys}
\begin{equation}
    S_n(M) \equiv \int_{s_0^-}^{s_0^+} ds \, s^n \, \Delta \Pi(s) \, {\rm e}^{-s^2/M^2} \: .
\end{equation}
The odd and even OPE, $o = S_0(M)$ and $e = - S_1(M)$, 
and their derivatives with respect to $M^{-2}$, 
$o^\prime = - S_3(M)$ and $e^\prime = S_4(M)$, 
can then be related via \eqref{eq:m_sys} to these moments. 
Thereby, new quantities $\overline{\Delta m}$ and $\overline{m}$ 
may be defined which encode the combined mass-width properties 
of the particles under consideration:
\begin{subequations}
\begin{align}
    \overline{\Delta m} \equiv & \frac12 \frac{S_1 S_2 - S_0 S_3}{S_1^2 - S_0 S_2} \: ,
    \\
    \overline{m_+ m_-} \equiv & - \frac{S_2^2 - S_1 S_3}{S_1^2 - S_0 S_2} \: 
\end{align}
\end{subequations}
and $\overline m^{\,2} \equiv \overline{\Delta m}^{\,2} + \overline{m_+ m_-}$.
For the above pole ansatz, these quantities become 
$\Delta m = \overline{\Delta m}$ and $m = \overline{m}$, i.e.,
they allow for an interpretation as mass splitting and mass centroid.
The relations (3.6) and (3.7) avoid the use of a special ansatz of the
spectral function, but prevent a direct physical and obvious interpretation.  
 
%================================================
\section{Evaluation for $D$ and $\overline D$ mesons}
\label{sct:evaluation_for_d_dbar_mesons}

We proceed with the above pole ansatz and evaluate the behavior 
of $m_\pm$ having in mind that these parameters characterize the combined $D, 
\overline D$ spectral functions, but need not necessarily describe the pole positions 
in general. According to the above defined current operators, 
$D$ stands either for $D^+$ or $D^0$ and $\overline{D}$ for $D^-$ or $\overline{D}^{\,0}$.

Because $dm_\pm/dM = 0$ has been used to derive \eqref{eq:m_sys} 
we have to look for the extrema of $m_\pm (M)$. Furthermore, in order to solve consistently 
the system of equations defined by \eqref{eq:mom_sys}, 
the values taken for $m_\pm$ must be fixed at the same Borel mass $M$. 
Therefore, we evaluate the sum rules using two threshold parameters 
$(s_0^\pm)^2 = s_0^2 \pm \Delta s_0^2$ and demand that the minima of the respective Borel curves $m_+(M)$ and $m_- (M)$ must be at a common Borel mass $M$. Hence, the thresholds are prescribed and offer the possibility to give a consistent solution to \eqref{eq:mom_sys}.

Analog to the analysis in \cite{Hayashigaki}, we chose the threshold parameter $s_0^2 = 6.0 \, \text{GeV}^2$, which approximately reproduces the vacuum case. At zero density we obtain for $m_\pm$ a value of $1.863$ GeV, representing a reasonable reproduction of the experimental value of the $D$ mass. The employed condensate values are listed in Tab.~\ref{tb:cond}. 

\begin{table}
\caption{\it \label{tb:cond} List of employed condensate parameters. A discussion of these numerical values can be found in \cite{Furnstahl}; further remarks on $\langle q^\dagger g \sigma \mathrsfs{G} q \rangle$ are given in \cite{Morath}. 
For the strong coupling we utilize $\alpha_s = 4 \pi / \left[ ((11 - 2 N_f/3) \ln(\mu^2/\Lambda_{QCD}^2)) \right]$ with $\mu$ being the renormalization scale, taken to be of the order of the largest quark mass in the system, and $N_f$ being the number of quark flavors with mass smaller than $\mu$; $\Lambda_{QCD}^2 = 0.25 \text{ GeV}^{\, 2}$ is the dimensional QCD parameter. The employed quark pole masses are $m_c = 1.5$ GeV and $m_b = 4.7$ GeV \cite{Narison2}.
}
    \begin{tabular}{|c|c|c|}
      \hline
      condensate & vacuum value $\langle \cdots \rangle_{vac}$ & density dependent part $\langle \cdots \rangle_{med}$ \\ \hline
      \hline
      $\langle \overline{q} q \rangle$ & $(-0.245 \text{ GeV})^3$ & $45/11 \, n$ \\
      \hline
      $\GGVren$ & $(0.33 \text{ GeV})^4$ & $- 0.65 \text{ GeV} \, n$ \\
      \hline
      $\langle \overline{q} g \sigma \mathrsfs{G} q \rangle$
        & $0.8 \text{ GeV}^2 \times (-0.245 \text{ GeV})^3$ & $3 \, n \text{ GeV}^2$ \\
      \hline
      $\langle q^\dagger q \rangle$ & $0$ & $1.5 \, n$ \\
      \hline
      $\GGMren$ & $0$ & $- 0.05 \text{ GeV} \, n$ \\
      \hline
      $\langle q^\dagger i D_0 q \rangle$ & $0$ & $0.18 \text{ GeV} \, n$ \\
      \hline
      $\langle \overline{q} \left[ D_0^2 - \frac18 g \sigma \mathrsfs{G} \right] q \rangle$ & $0$ & $-0.3 \text{ GeV}^2 \, n$ \\
      \hline
      $\langle q^\dagger D_0^2 q \rangle$ & $0$ & $-0.0035 \text{ GeV}^2 \, n$ \\
      \hline
      $\langle q^\dagger g \sigma \mathrsfs{G} q \rangle$ & $0$ & $0.33 \text{ GeV}^2 \, n$ \\
      \hline
    \end{tabular}
\end{table}

\begin{figure}
    \centering
    		%ps:width=0.35, pdf:width=0.49
        %\rotatebox{270}{
        	\includegraphics[width=0.49\textwidth]{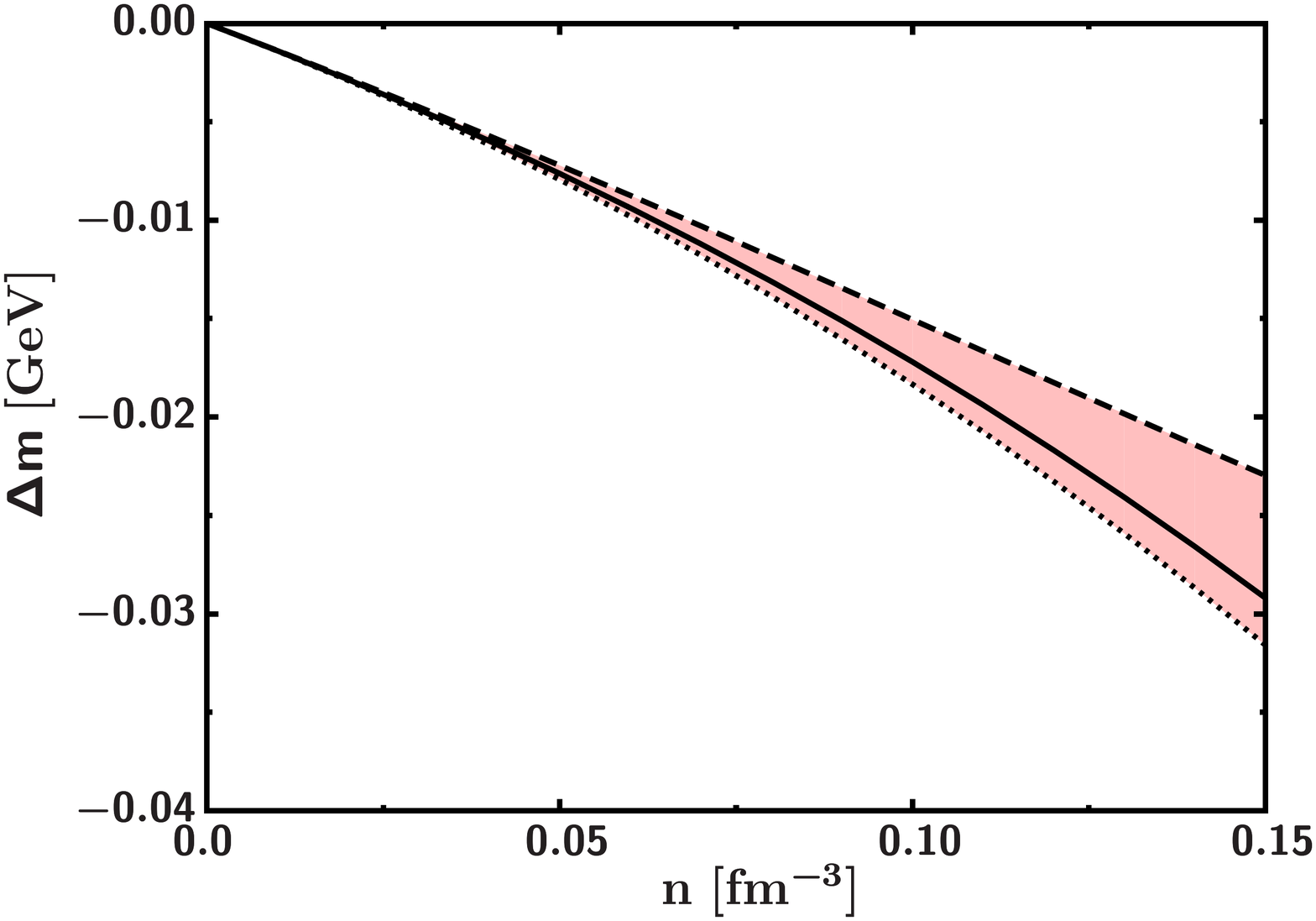}
        %}
        %\rotatebox{270}{
        	\includegraphics[width=0.49\textwidth]{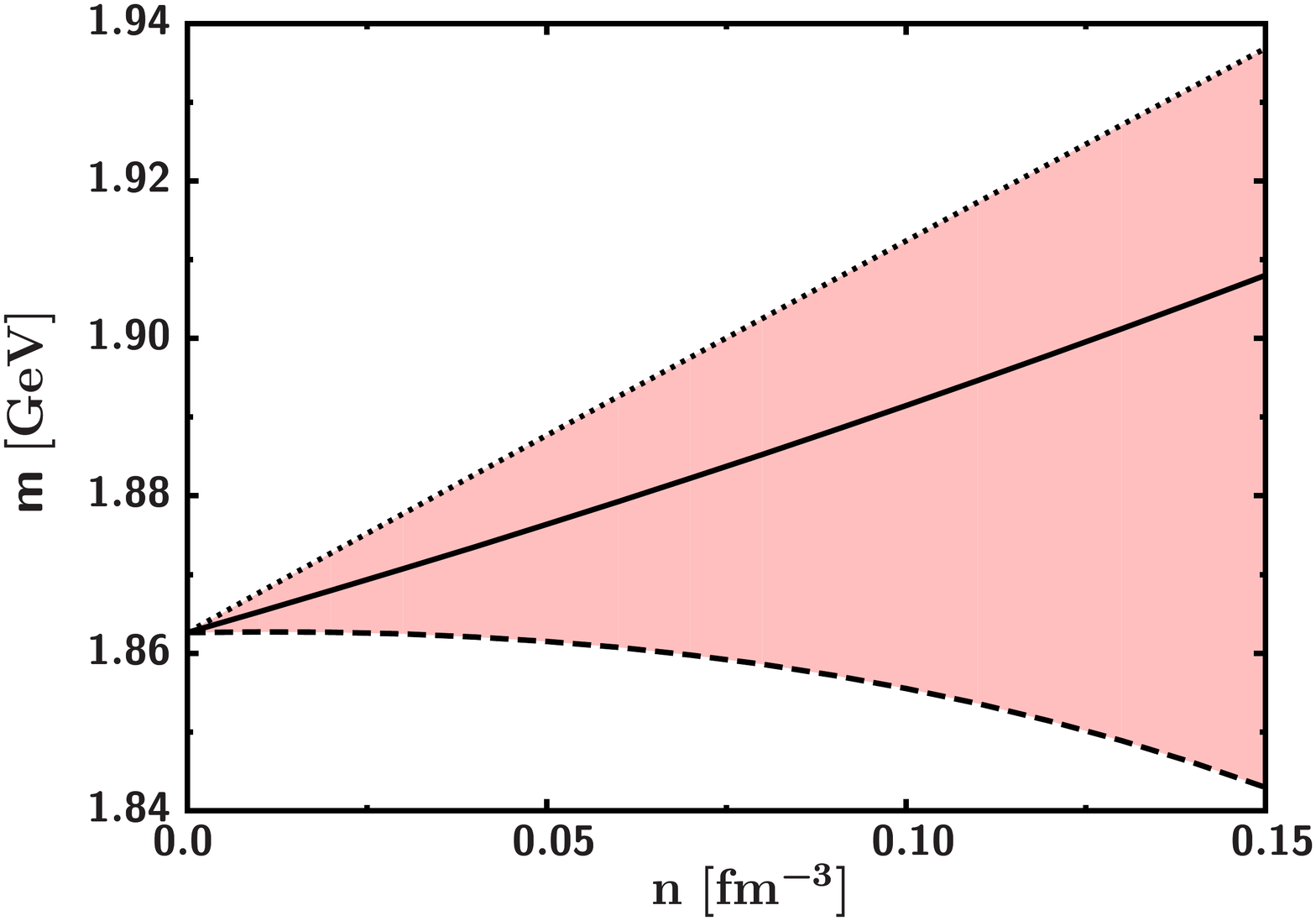}
        %}
    \caption{\it Mass splitting parameter $\Delta m$ (left) and mass centroid $m$ (right) for $D$, $\overline D$ mesons for density independent threshold (solid line) and a density dependent threshold $s_0^2(n) = s_0^2(0) \pm n/n_0 \text{ GeV}^{\, 2}$, where the dotted (dashed) curve is for the positive (negative) sign. Note that the mass splitting in the iso-doublet is $2 \Delta m$.}
    \label{fig:D_sn}
\end{figure}

The density dependence of the mass splitting parameter $\Delta m$ and the $D + \overline D$ doublet mass centroid $m$ are exhibited in Fig.~\ref{fig:D_sn} as a function of the density. We observe an almost linear behavior of the mass splitting with increasing density. At $n = 0.15 \text{ fm}^{-3}$ a mass splitting of $2 \Delta m \approx - 60 \, \text{MeV}$ is obtained. The mass splitting has negative values, i.e. $m_- > m_+$ or $m_{\overline D} > m_D$ in line with previous estimates in \cite{Morath}. For the mass centroid $m$ our result differs from the one in \cite{Hayashigaki}, where a mass shift of the order of $-50$ MeV is obtained, while we find about $+45$ MeV. At $n = 0.15 \text{ fm}^{-3}$ the splitting of the threshold parameters is $\Delta s^2_0 \approx - 0.3 \text{ GeV}^2$ for the used set of parameters, and the minima of the Borel curves are located at $M \approx 0.95$ GeV and are slightly shifted upwards with increasing density.

While the mass splitting is fairly robust, we find a sensitivity of the centroid mass shift under variation of the continuum threshold parameter $s_0^2$. The above reported value of the mass centroid changes towards zero when lowering $s_0^2$. In Fig.\ \ref{fig:D_sn} we therefore also use a density dependent prescription for the threshold $s_0^2(n) = s_0^2(0) \pm n/n_0 \text{ GeV}^2$, where $n_0 = 0.15 \text{ fm}^{-3}$ is the nuclear saturation density; $\pm 1/n_0 \text{ GeV}^2$ corresponds to the first Taylor coefficient $ds_0^2/dn(0)$. This simple choice enables us to identify the uncertainties which might emerge due to the introduction of a density independent threshold.
As can be seen, the average mass shift may change in sign. In contrast, the result for $\Delta m$ shows only a weak dependence on $s_0^2$.

At this point a comment concerning the sign of $\langle q^\dagger g \sigma \mathrsfs{G} q \rangle$ is order. If one would use $\langle q^\dagger g \sigma \mathrsfs{G} q \rangle = -0.33 \text{ GeV}^2 \, n$ instead (this option is also discussed in \cite{Furnstahl}, $\langle q^\dagger D_0^2 q \rangle$ would acquire a value of $-0.0585 \text{ GeV}^2 \, n$ accordingly) one would get a much larger mass splitting of about $-180$ MeV, which is far beyond the estimates obtained in \cite{Tolos,Angel,Lutz_charm}. Hence, we favor the positive sign of $\langle q^\dagger g \sigma \mathrsfs{G} q \rangle$ as advocated in \cite{Morath}, too. Clearly, further correlators should be studied to investigate the role of the condensate $\langle q^\dagger g \sigma \mathrsfs{G} q \rangle$.

We emphasize the special evaluation strategy employed so far. Other possibilities are, e.g., variation of $s_0^2$ and $\Delta s^2$ so that $m_\pm(M)$ develop a section of maximum flatness. Interestingly, this method leads to a rather low threshold $s_0^2 \approx 4 \text{ GeV}^2$ and a low vacuum mass of about $m \approx 1.6$ GeV.
In contrast, averaging over the Borel curves in the interval $[0.9 M_0, 1.2 M_0]$, around the minimum $M_0$, we find the values for the mass splitting $\Delta m \approx -40 \text{ MeV}$ and the average mass shift to be of the same order as quoted above, whereas the absolute value of the vacuum mass becomes $m = 1.877$ GeV.

\begin{figure}
    \centering
    		%ps:width=0.35, pdf:width=0.49
        %\rotatebox{270}{
        	\includegraphics[width=0.49\textwidth]{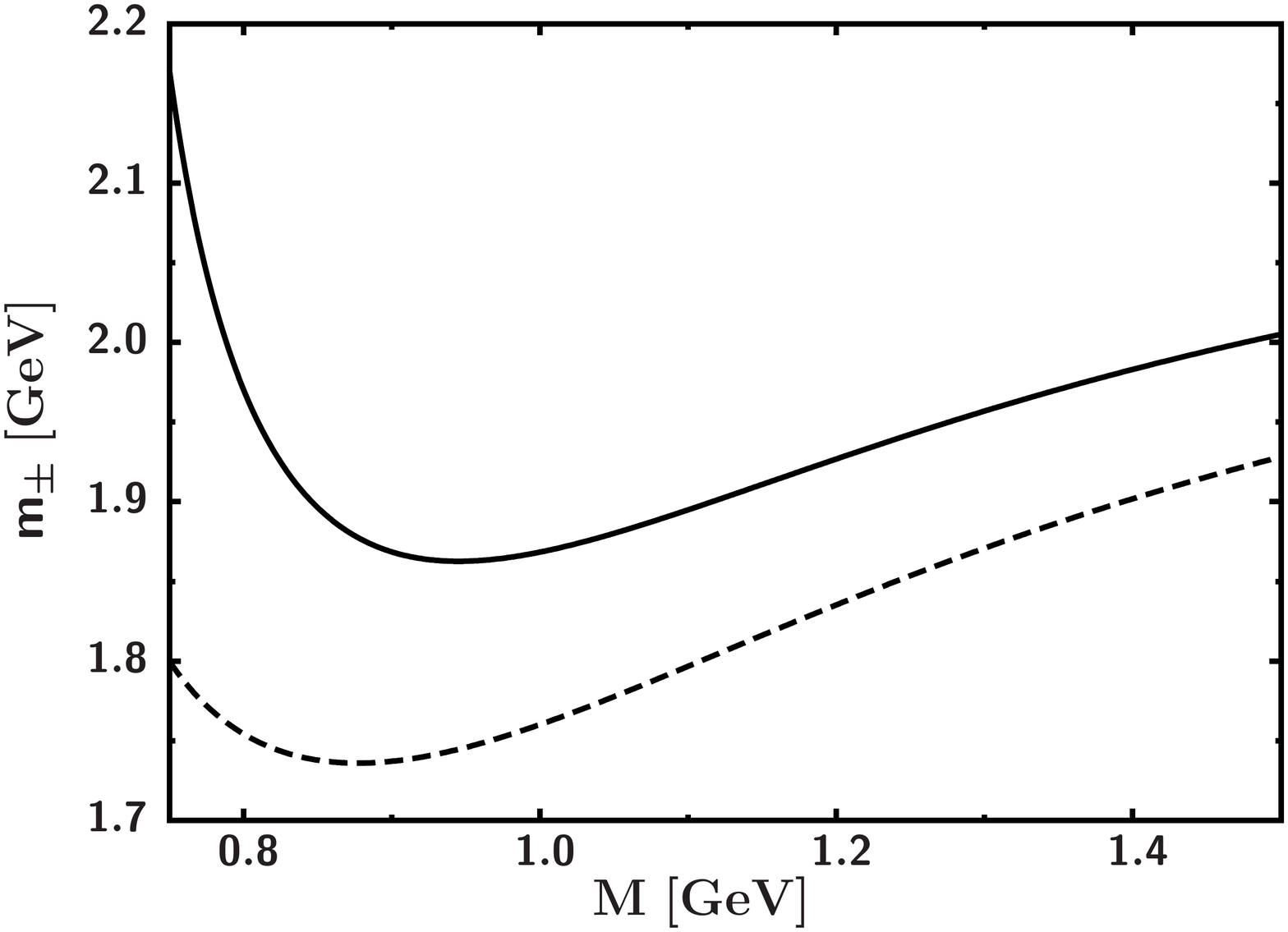}
        %}
        %\rotatebox{270}{
        	\includegraphics[width=0.49\textwidth]{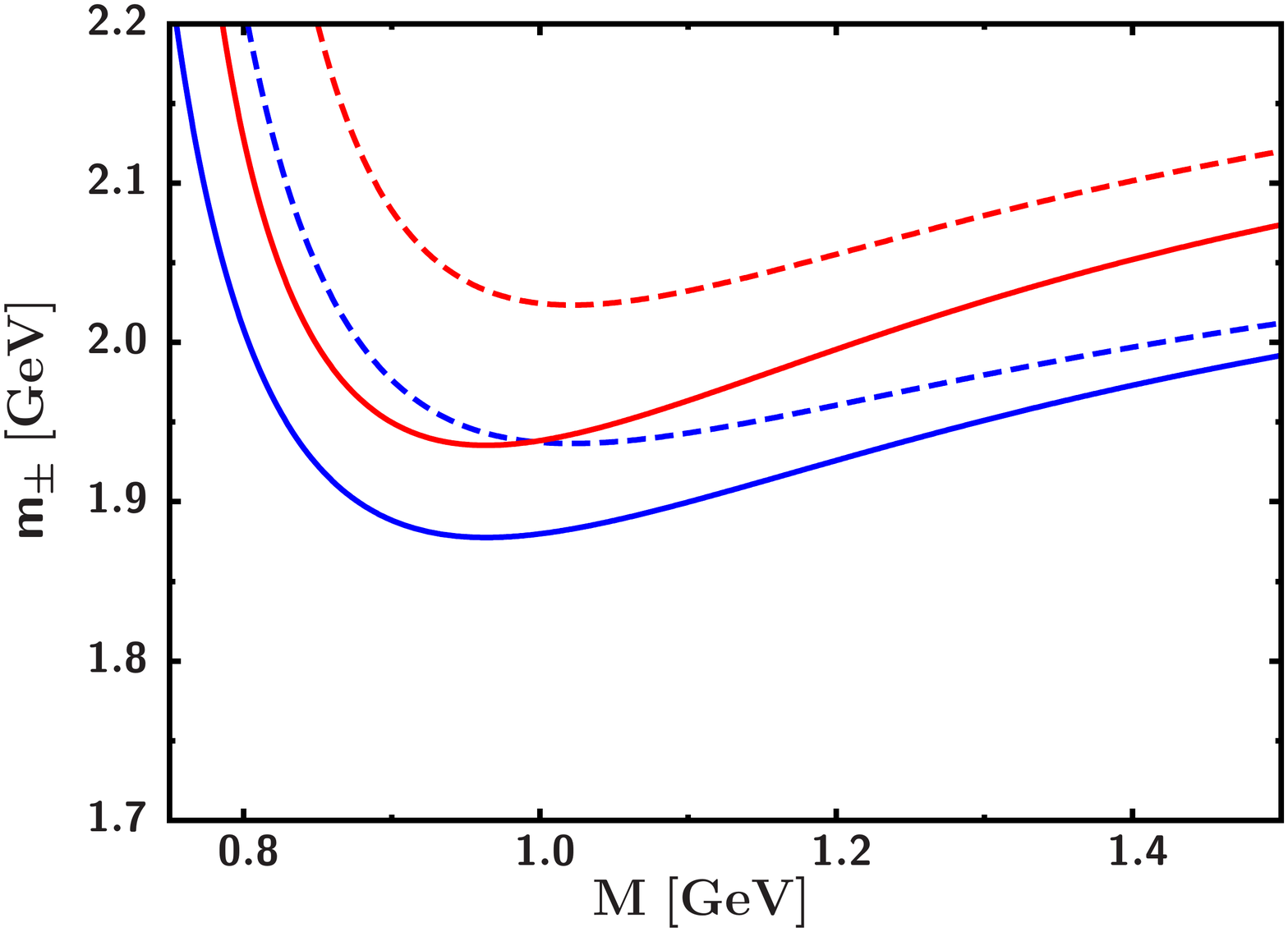}
        %}
    \caption{\it Borel curves $m_\pm (M)$ for the D meson for two densities and two values of the chiral condensate.
        Left panel: $n=0$,
        right panel: $n=0.15 \, \text{fm}^{-3}$. (Solid curves: chiral condensate from Tab.~\ref{tb:cond},
        dotted curves: doubling the chiral condensate (left panel) or doubling the density dependent part of the chiral condensate (right panel);
        lower (upper) curves in the right panel are for $m_+$ ($m_-$), while $m_+ = m_-$ for the vacuum case in the left panel.).
        }
    \label{fig:vary_qq}
\end{figure}

Let us now further consider the impact of various condensates. The result for the mass splitting $\Delta m$ strongly depends on the quark density $\langle q^\dagger q \rangle$, whose density dependence is uniquely fixed. The odd mixed quark-gluon condensate $\langle q^\dagger g \sigma G q \rangle$ and the chiral condensate $\langle \overline{q} q \rangle$ are the next influential ones for the mass splitting.
The density dependent part of the chiral condensate enters in order ${\cal O}(n^2)$ gaining its influence from the heavy quark mass amplification factor. The influence of the chiral condensate is illustrated in Fig.~\ref{fig:vary_qq}. In a strictly linearized sum rule evaluation, the density dependent part of $m_c \langle \overline q q \rangle$ would be omitted for the mass splitting.
However, numerically the influence of the chiral condensate is of the same order as (but still smaller than) the above discussed condensate $\langle q^\dagger g \sigma \mathrsfs{G} q \rangle$, which enters the odd part of the OPE. As expected, the density dependence of the mass centroid is basically determined by the even part of the OPE.
The density dependent parts of the other even condensates are of minor importance for the mass splitting. The shift of the centroid's mass is anyhow fragile.

\begin{figure}
    \centering
    		%ps:width=0.35, pdf:width=0.49
        %\rotatebox{270}{
        	\includegraphics[width=0.49\textwidth]{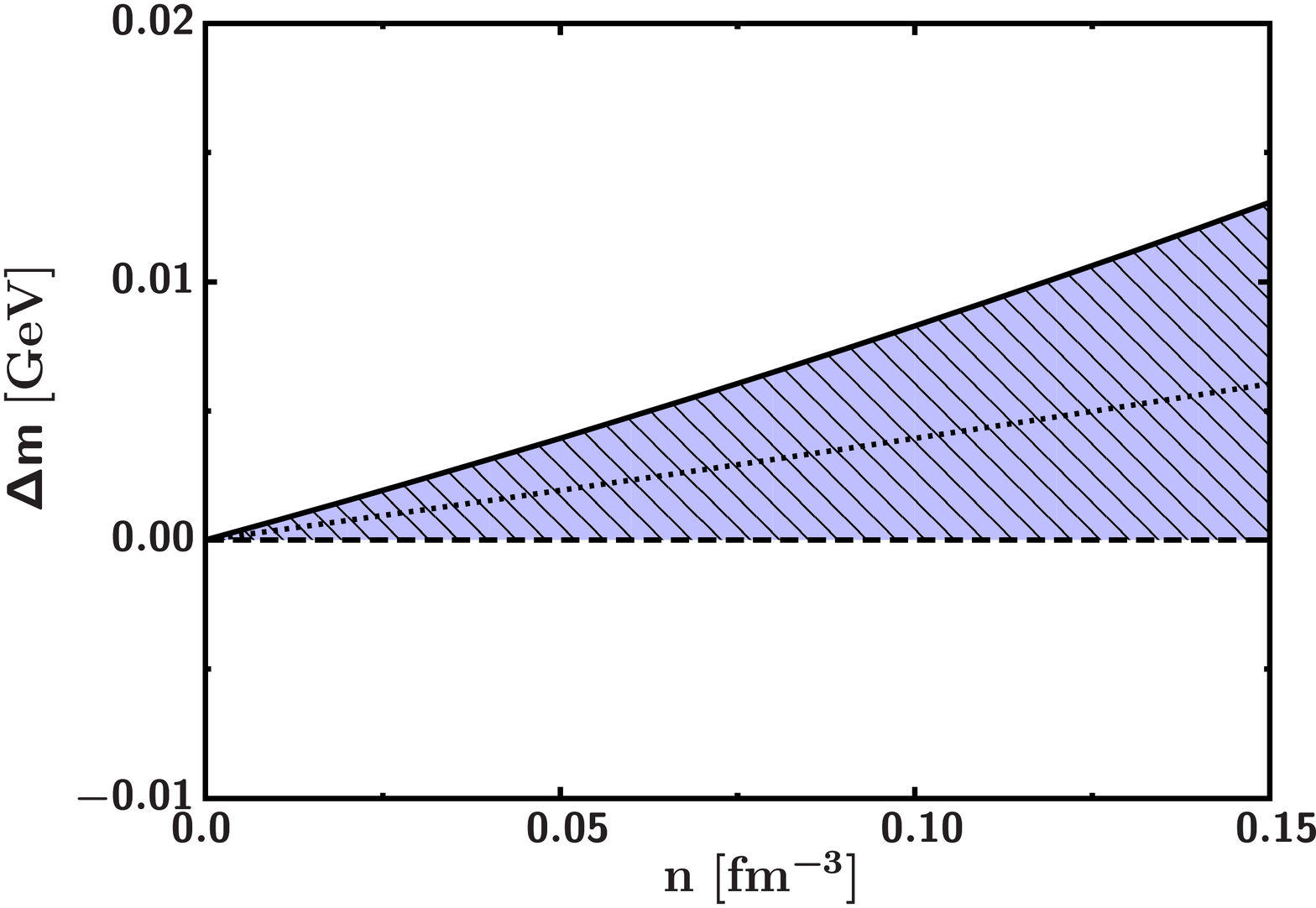}
        %}
        %\rotatebox{270}{
        	\includegraphics[width=0.49\textwidth]{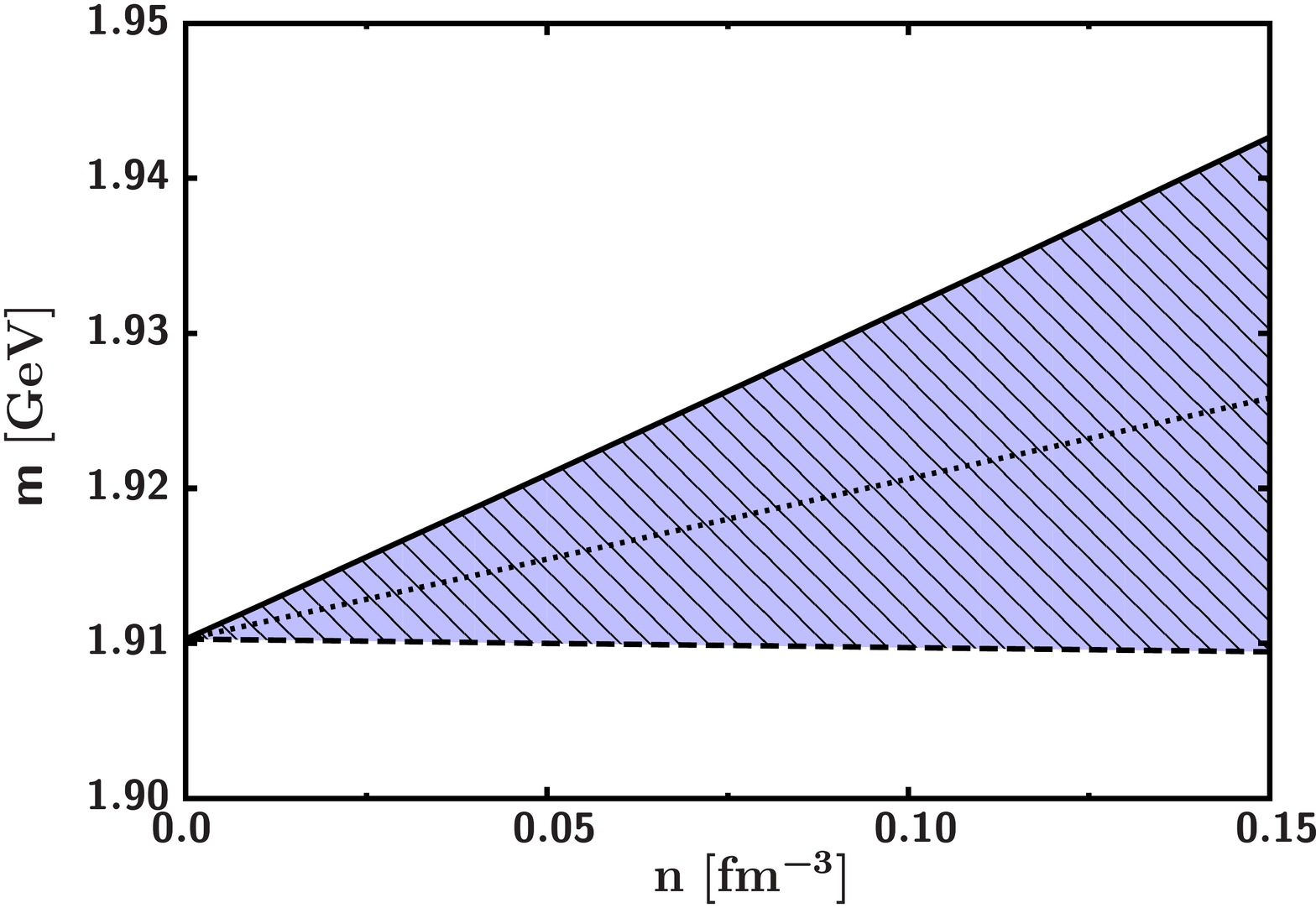}
        %}
    \caption{\it $\Delta m$ (left) and $m$ (right) for $D_s$ and $\overline D_s$ at $s_0^2 = 8.0 \text{ GeV}^{\,2}$ and for $y=0.5$ (solid), $y=0.25$ (dotted), $y=0$ (dashed).}
    \label{fig:m_delta_Ds}
\end{figure}

Within the given formulation and with the first evaluation strategy, one may also consider $D_s$ and
$\overline D_s$ mesons with the replacements $m_q \to m_s$,
$\langle \overline q q \rangle \to \langle \overline s s\rangle = 0.8 \langle \overline q q \rangle_{vac} + y \langle \overline q q \rangle_{med}$,
$\langle \overline{q} g \sigma \mathrsfs{G} q \rangle \to \langle \overline{s} g \sigma \mathrsfs{G} s \rangle = 0.8 \, \text{GeV}^2 \, \langle \overline{s} s \rangle$,
$\langle q^\dagger q \rangle \to \langle s^\dagger s \rangle = 0$,
$\langle q^\dagger i D_0 q \rangle \to \langle s^\dagger i D_0 s \rangle = 0.018 \, \text{GeV} \, n$,
$\langle \overline{q} \left[ D_0^2 - \frac18 g \sigma \mathrsfs{G} \right] q \rangle \to \langle \overline{s} \left[ D_0^2 - \frac18 g \sigma \mathrsfs{G} \right] s \rangle = y \langle \overline{q} \left[ D_0^2 - \frac18 g \sigma \mathrsfs{G} \right] q \rangle$,
$\langle q^\dagger D_0^2 q \rangle \to \langle s^\dagger D_0^2 s \rangle = y \langle q^\dagger D_0^2 q \rangle$,
$\langle q^\dagger g \sigma \mathrsfs{G} q \rangle \to \langle s^\dagger g \sigma \mathrsfs{G} s \rangle = y \langle q^\dagger g \sigma \mathrsfs{G} q \rangle$.
The anomalous strangeness content of the nucleon is varied as $0 \leq y \leq 0.5$ \cite{Navarra}; lattice calculations, for example, point to $y=0.36$ \cite{Dong}.
The results are exhibited in Fig.~\ref{fig:m_delta_Ds}. At $n = 0.15 \text{ fm}^{-3}$ and $y=0.5$ we observe a mass splitting of $2\Delta m \approx +25$ MeV and a shift of the mass centroid of about $+30$ MeV. The splitting of the thresholds becomes $\Delta s_0^2 \approx 0.83 \text{ GeV}^{\,2}$, and the minima of the Borel curves are located at $M \approx 0.89$ GeV and slightly shifted upwards with increasing density.
The main reason for the positive sign of the mass splitting is the vanishing strange quark net density $\langle s^\dagger s \rangle$. The mass splitting acquires positive values for $\langle s^\dagger s \rangle \lesssim 0.4 \, n$ (at $y=0.5$).
Mass splitting and the average mass shift tend to zero for $y \to 0$. In this case only the pure gluonic condensates, which enter the even OPE and are numerically suppressed compared to other condensates, have a density dependence.
Note that these evaluations are, at best, for a rough orientation, as mass terms $\propto m_s$ have been neglected. The too low vacuum mass of $1.91$ GeV compared to the experimental value $m_{D_s} = 1.968$ GeV is an indication for some importance of strange quark mass terms. Such mass terms $\propto m_s$ have been accounted for in \cite{Hayashigaki_Terasaki} for the vacuum case. The complete in-medium OPE and sum rule evaluation deserves separate investigations, as $m_s$ introduces a second mass scale.

%================================================
\section{Evaluation for $B$ and $\overline B$ mesons}
\label{sct:evaluation_for_b_bbar_mesons}

\begin{figure}
    \centering
    		%ps:width=0.35, pdf:width=0.49
        %\rotatebox{270}{
        	\includegraphics[width=0.49\textwidth]{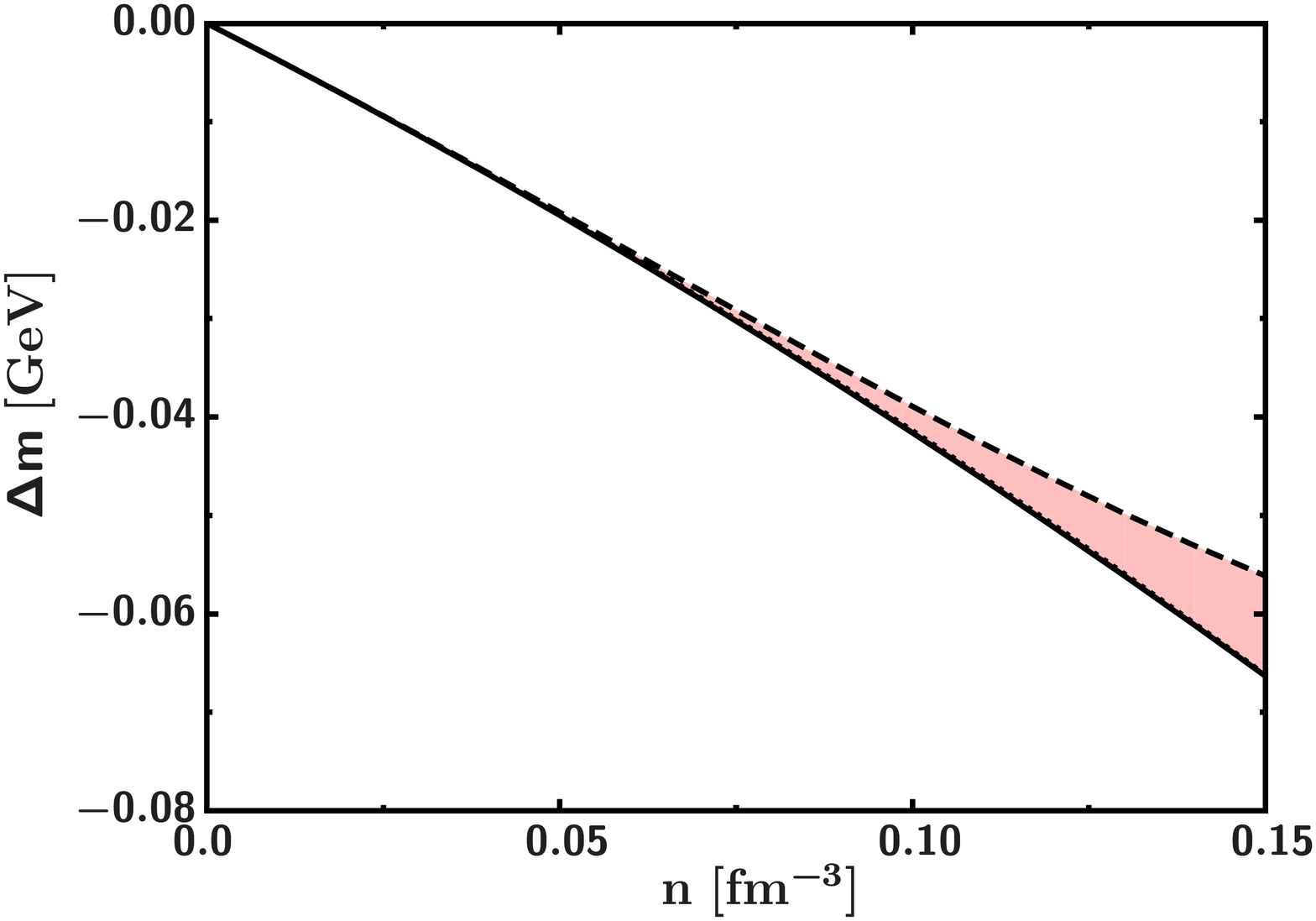}
        %}
        %\rotatebox{270}{
        	\includegraphics[width=0.49\textwidth]{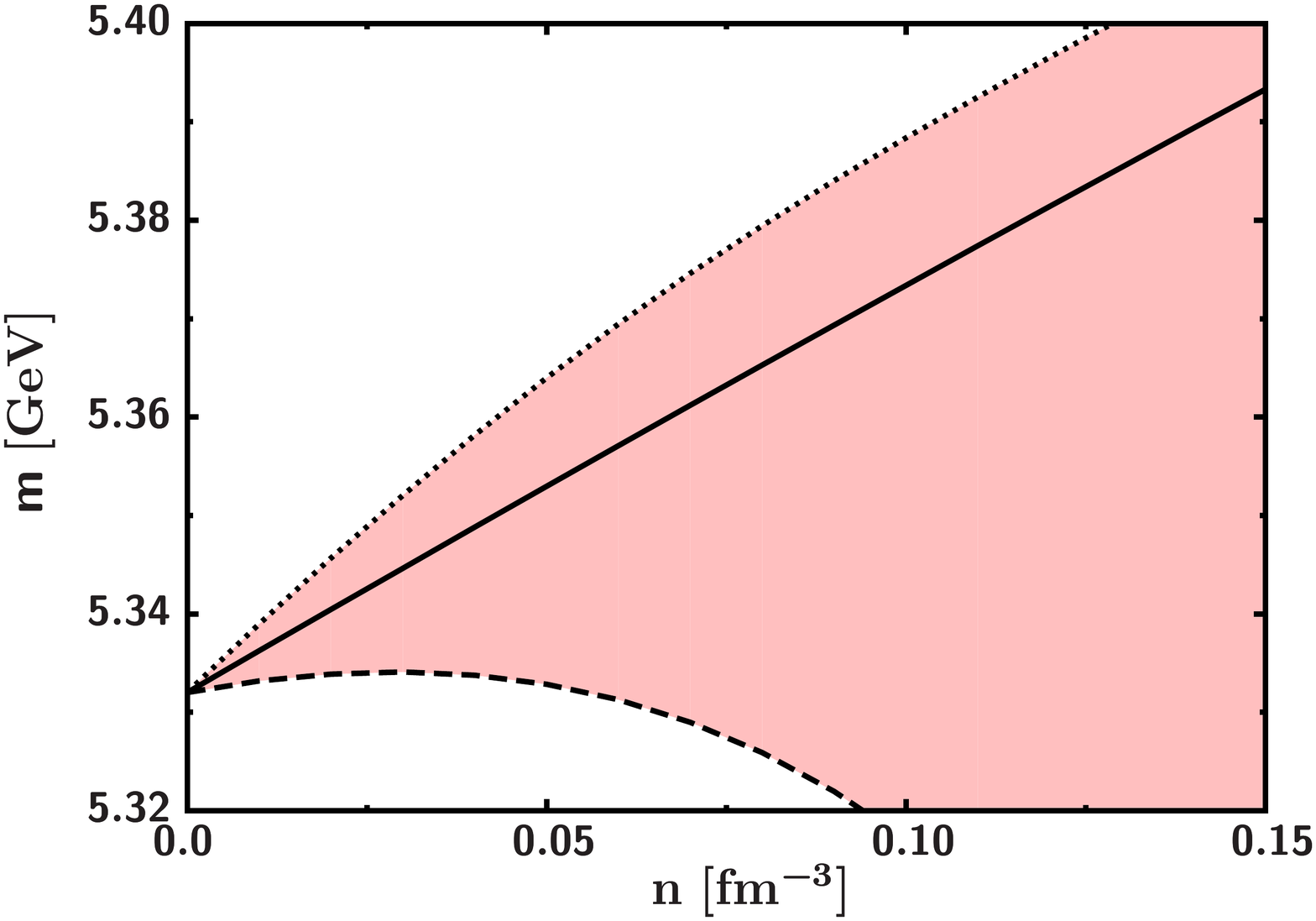}
        %}
    \caption{\it $\Delta m$ (left) and $m$ (right) for $\overline B$ and $B$ at $s_0^2 = 40  \text{ GeV}^{\,2}$ and $m_b = 4.7$ GeV. For line codes see Fig.~\ref{fig:D_sn}. For density dependent thresholds, $s_0^2 = s_0^2(0) \pm 7 n / n_0 \, \text{GeV}^{\,2}$ has been used.}
    \label{fig:m_delta_B}
\end{figure}

We turn now to $B$ and $\overline B$ mesons. The corresponding current operators are $\text{j}_{B^+} = i \overline{b} \gamma_5 u$ or $\text{j}_{B^0} = i \overline{b} \gamma_5 d$. The antiparticles correspond to $\text{j}_{B^-} = \text{j}^\dagger_{B^+} = i \overline{u} \gamma_5 b$ or $\text{j}_{\overline{B}^0} = \text{j}^\dagger_{B^0} = i \overline{d} \gamma_5 b$. The above equations and, in particular, the OPE are applied with the replacements $m_c \to m_b$ and $m_{B^\pm} \to m_{\mp}$ in order to take into account the distinct heavy-light structure compared to the $D$ meson case.
The Borel curves $m_\pm (M)$ display, analog to the case of open charm, pronounced minima at a Borel mass of about $1.7$ GeV. We utilize again the first evaluation strategy. Numerical results are exhibited in Fig.~\ref{fig:m_delta_B}. We employ $s_0^2 = 40 \text{ GeV}^{\, 2}$ and obtain $m \approx 5.33$ GeV for the vacuum mass. One observes a mass splitting of $2\Delta m \approx - 130 \text{ MeV}$ at $n = 0.15 \text{ fm}^{-3}$. The centroid is shifted upwards by about $60$ MeV. The splitting of the threshold parameters becomes $\Delta s^2_0 \approx -3.4 \text{ GeV}^2$ and the minima of the Borel curves $m_\pm(M)$ are shifted from $M \approx 1.67$ GeV in vacuum to $M \approx 1.71$ GeV at $n = 0.15 \text{ fm}^{-3}$.
In case of $\overline B, B$ mesons, the combination $m_b \langle \overline d d \rangle$ is expected to have numerically an even stronger impact than the term $m_c \langle \overline d d \rangle$ in the charm sector. Indeed, the influence of the chiral condensate becomes even larger than that of the odd mixed quark-gluon condensate $\langle q^\dagger g \sigma \mathrsfs{G} q \rangle$ at higher densities. 
The overall pattern resembles the results exhibited in Fig.~\ref{fig:vary_qq}, but with shifted mass scale for $m$. The other evaluation strategies yield the same results. Setting $\langle q^\dagger g \sigma \mathrsfs{G} q \rangle = -0.33 \text{ GeV}^2 \, n$, and, hence, $\langle q^\dagger D_0^2 q \rangle = -0.0585 \text{ GeV}^2 \, n$, a mass splitting of $2 \Delta m \approx - 220$ MeV and an average mass shift $\approx 45$ MeV would be obtained.

%================================================
\section{Summary}
\label{sct:summary}

In summary we have evaluated the Borel transformed QCD sum rules for pseudo-scalar mesons composed of a combination of a light and a heavy quark.
The heavy quark mass introduces a new scale compared to QCD sum rules in the light quark sector.
The evaluation of the sum rules, complete up to mass dimension 5, has been performed for $D, \overline D$ and $\overline B, B$ mesons with a glimpse on $D_s, \overline D_s$ as well. Our analysis relies on the often employed pole + continuum ansatz for the hadronic spectral function.
This is a severe restriction of the generality of the practical use of sum rules.
In this respect, the extracted parameters refer to this special ansatz and should be considered as indicators for changes of the true spectral functions of hadrons embedded in cold nuclear matter.
Particles and antiparticles are coupled -- a problem which is faced also for hadrons with conserved quantum numbers in the light
quark sector \cite{Erice,Furnstahl,our NPA}.

We presented a transparent approximation to highlight the role of the even and odd parts of the OPE.
Numerically, we find fairly robust mass splittings (for the employed set of condensate values) in the iso-doublets, while an assignment of a possible mass shift of the centroids is not yet on firm ground.
The impact of various condensates is discussed, and $\langle q^\dagger q \rangle$, $\langle q^\dagger g \sigma \mathrsfs{G} q \rangle$ and $\langle \overline q q \rangle$ are identified to drive essentially the mass splitting. While $\langle \overline q q \rangle$ is amplified by the heavy quark mass, it enters nevertheless the sum rules beyond the linear density dependence. A concern is the sign of $\langle q^\dagger g \sigma \mathrsfs{G} q \rangle$, vanishing in vacuum, which determines the size of the $D - \overline D$ mass splitting.
These findings, in particular for $D, \overline D$, $D_s, \overline D_s$, are of relevance for the planned experiments at FAIR.

Acknowledgments:
The authors gratefully acknowledge discussions with 
S. Leupold, M. Lutz, W. Weise and S. Zschocke.
The work is supported by BMBF 06DR136, GSI-FE and EU I3HP.

%================================================

\end{document}